\documentclass[10pt,conference]{seip/IEEEtran}
\usepackage{tikz}
\usepackage{fontawesome5}
\usepackage{placeins}
\usepackage{array}
\usepackage{tabularx}
\usepackage{multirow}
\usepackage{fvextra}
\usepackage{soul}
\usepackage{booktabs}
\usepackage{url}
\usepackage{cite}
\usepackage{enumitem}
\usepackage[skins]{tcolorbox}
\usetikzlibrary{arrows.meta,positioning}
\newcommand{\system}{RefactorForge}
\newcounter{pipelinestep}
\newcommand{\pipelinestep}[1]{
  \refstepcounter{pipelinestep}
  \subsubsection*{Step \thepipelinestep. #1}
}
\newcommand{\baselinemodel}{\texttt{Baseline-\allowbreak92B}}
\newcommand{\refactoringmodel}{\texttt{Refact-\allowbreak92B}}
\newcommand{\onboardingAssessed}{729}
\newcommand{\onboardingBuilt}{614}
\newcommand{\onboardingTested}{476}
\newcommand{\onboardingAccepted}{306}

\newcommand{\onboardingBuildRate}{84.2\%}
\newcommand{\onboardingTestExecutionRate}{65.3\%}
\newcommand{\onboardingAcceptanceRate}{42.0\%}
\newcommand{\onboardingLLMSpend}{\$11.38}

\newcommand{\onboardingCentsPerAccepted}{3.72}

\newcommand{\rqonePromaxAbsoluteGain}{7.06}

\newcommand{\rqoneRefactorbenchAbsoluteGain}{18.00}

\newcommand{\rqtwoVerifiedAbsoluteGain}{4.97}

\newcommand{\rqtwoProAbsoluteGain}{1.23}

\providecommand{\rqthreeVerifiedMiniPValue}{\ensuremath{p=\mathrm{0.002}}}

\newtcolorbox{summarybox}[1][Summary]{
  enhanced,
  title=#1,
  fonttitle=\bfseries\small,
  fontupper=\small,
  colback=gray!6,
  colframe=gray!70!black,
  coltitle=white,
  boxrule=0.5pt,
  arc=1pt,
  left=4pt, right=4pt, top=3pt, bottom=3pt,
  before skip=6pt, after skip=6pt
}
\newcommand{\Description}[1]{}
\setlist[description]{style=nextline,leftmargin=0pt,labelindent=0pt}

\renewenvironment{figure}{\begin{figure*}[t]}{\end{figure*}}
\renewenvironment{table}{\begin{table*}[t]}{\end{table*}}
\begin{document}
\title{Software Engineering LLMs: Why We Need Them and How Software Engineering Helps Build Them}
\author{
\IEEEauthorblockN{
Amir M. Ebrahimi\IEEEauthorrefmark{1},
Mohammed Mehedi Hasan\IEEEauthorrefmark{1},
Seyed Shayan Daneshvar\IEEEauthorrefmark{2},\\
Jiahuei Lin\IEEEauthorrefmark{3},
Gopi Krishnan Rajbahadur\IEEEauthorrefmark{1},
Ahmed E. Hassan\IEEEauthorrefmark{1}}
\IEEEauthorblockA{
\IEEEauthorrefmark{1}Queen's University, Canada;\quad
\IEEEauthorrefmark{2}University of Manitoba, Canada\\
\IEEEauthorrefmark{3}Center for Software Excellence, Huawei Technologies, Canada\\
amir.ebrahimi@queensu.ca\quad
mohammedmehedi.hasan@queensu.ca\quad
daneshvs@myumanitoba.ca\\
jhlin@cs.queensu.ca\quad
grajbahadur@acm.org\quad
ahmed@cs.queensu.ca}
}
\maketitle
\begin{abstract}

Software engineering (SE) expertise can shape how coding large language
models (LLMs) are trained and which engineering capabilities they acquire.
We examine this role through an industrial case study of capability-focused
training for a 92B-parameter coding LLM serving more than 100,000 developers.
We demonstrate how established SE knowledge can guide the selection of
engineering capabilities, the construction of verifiable training data,
and the evaluation of model improvements. To target repository-level
refactoring, we developed \system{}, a pipeline that combines refactoring
knowledge, established SE tools, and verification techniques to construct
training data. Across \onboardingAccepted{} Python repositories, \system{}
generated 5,704 executable tasks and retained 4,572 verified teacher
trajectories totaling 72.51 million tokens. Incremental supervised
fine-tuning on these trajectories, accounting for 0.46\% of the preceding
coding mixture's tokens, improved RefactorBench performance by \rqoneRefactorbenchAbsoluteGain{} percentage
points and SWE-Bench ProMax performance by \rqonePromaxAbsoluteGain{} points.
Under OpenCode, these improvements extended to SWE issue resolution,
with gains of \rqtwoVerifiedAbsoluteGain{} points on SWE-Bench Verified
and \rqtwoProAbsoluteGain{} points on SWE-Bench Pro. Improvements also
persisted across other coding-agent harnesses, while greenfield repository
generation showed mixed results. We report lessons from constructing
capability-focused training data and evaluating its effects across
engineering workloads. Building on this industrial experience, we propose
a research agenda for expanding SE's role in transforming coding LLMs into
SE LLMs: incorporating SE knowledge into model training, applying engineering
methods to model development and maintenance, and systematically defining
the engineering capabilities these models should acquire.

\end{abstract}

\section{Introduction}
\label{sec:introduction}

\begin{seipregularfigure}[!t]
\centering
\includegraphics[width=0.80\columnwidth]{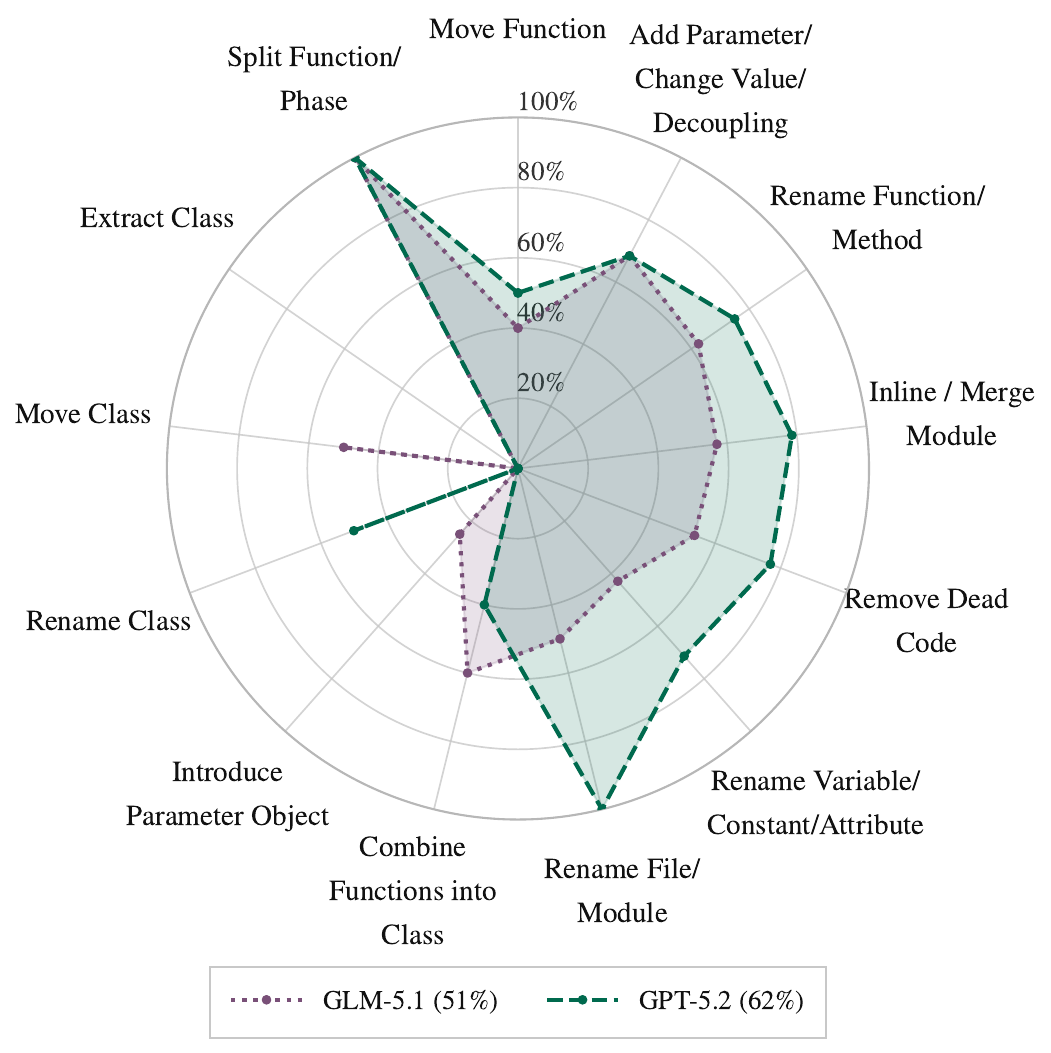}
\caption{Operation-level success rates of GPT-5.2 and GLM-5.1 across
13 RefactorBench categories. Aggregate success rates of 62\% and 51\% conceal
substantial differences across individual operations.}
\label{fig:frontier-radar}
\end{seipregularfigure}

Large language models (LLMs) and coding agents are reshaping software development in
industry and open source. Google reports that AI generates 75\% of its new
code~\cite{google2026cloudnext}, while Robbes et al.~\cite{robbes2026agentic} estimate
coding-agent adoption at 22 to 29\% across 128,018 GitHub projects. Frontier coding LLMs
now score above 96\% on SWE-Bench Verified: Claude Opus 5, DeepSeek V4 Pro, and GPT-5.6 Sol
reach 97.0\%, 96.4\%, and 96.2\%, respectively, under the same minimal agent
harness~\cite{valsai2026swebench}.

Strong coding performance, however, does not guarantee the engineering capabilities
required to maintain and evolve software. Whitfill et al.~\cite{whitfill2026swebenchmerge}
found that automated benchmark scores exceeded estimated maintainer acceptance rates by
24.2 percentage points, but code quality and compatibility were reported concerns.
Open-source evidence reveals problems after merge: across 182 repositories, Xia
and Miller~\cite{xia2026violent} found higher corrective-maintenance rates for agentic code
and associated higher rates of unreviewed merges with greater maintenance burden.
Software engineering requires more than completing difficult programming
tasks. Engineers must preserve compatibility, manage dependencies, remove unnecessary
code, and maintain systems as requirements evolve. Refactoring is one such software engineering capability that goes beyond code generation. Refactoring supports long-term maintainability by improving
software structure while preserving behavior
~\cite{fowlerRefactoringDefinition,kim2014refactoring}. Although
developers increasingly delegate refactoring to coding agents, which pair coding LLMs with
harnesses that supply tools and prompts~\cite{huang2025vibe}, frontier coding LLMs perform
unevenly across these
operations. To examine this gap, we evaluated GPT-5.2 and GLM-5.1 on
RefactorBench, which contains repository-level, multi-file refactoring tasks across 13
operation categories (see Section~\ref{sec:design} for the evaluation setup). The models achieved overall success rates of 62\% (GPT-5.2) and 51\% (GLM-5.1), but both failed every Extract Class task. GPT-5.2 also failed every Move Class and Introduce Parameter Object
(Figure~\ref{fig:frontier-radar}). These results reveal that even highly capable coding LLMs
struggle with established software engineering operations.

What coding LLMs learn depends on the training data developers can provide and the
improvements they reward. Pretraining develops broad programming knowledge, while
subsequent specialization and post-training strengthen selected capabilities through
curated data and feedback. Rajbahadur et al.~\cite{rajbahadur2026llm} describe
post-training as programming with data: training examples act as instructions, their
mixture forms the program, and the checkpoint is the compiled result. Repository
histories and executable tests make issue resolution an attractive training target, as
SWE-Gym and SWE-smith demonstrate~\cite{swegym,swesmith}. OpenAI, Anthropic, and
Microsoft describe evaluation-driven development, where measured weaknesses guide
subsequent improvements~\cite{openai2025evals,anthropic2026evals,mai_thinking_1}. Under
finite budgets, public leaderboard incentives can further steer training toward benchmark
gains, which Chen et al.~\cite{chen2026leaderboard} call \emph{benchmaxxing}. This
practice raises three challenges. First \textbf{(C1)}, data availability and evaluation priorities
can leave gaps in verified training data needed to improve specific engineering capabilities
and demonstrate those improvements~\cite{gorinova2026position}. \textbf{Second (C2)},
benchmark-oriented gains do not consistently transfer to other coding
tasks~\cite{shibaev2026don}. \textbf{Third (C3)}, training trajectories encode harness-specific
conventions, and improvements can diminish or reverse when the harness
changes~\cite{orchard,thangarajah2026dcas}. These challenges highlight the limits of relying on readily available training data and benchmark gains to develop engineering competence systematically.

A coding LLM serving diverse engineering workloads needs capabilities that extend beyond
the benchmarks and harnesses used to develop them. Our team maintains a coding LLM used by
over 100,000 developers across diverse repositories, workflows, and systems that have
evolved over decades (Section~\ref{sec:motivation}). Developers access the coding LLM
through different coding agents and rely on it to maintain existing software as well as
implement new functionality. We therefore explored a complementary hypothesis, which we
call \emph{capability-focused training}: \textbf{strengthening a foundational engineering
capability can improve the related tasks and workflows that depend on it.} We selected
refactoring as an initial case because it exercises code localization, dependency
reasoning, coordinated modification, and behavior preservation. The challenge was to
construct reliable training data for these operations and determine whether the
improvements transferred beyond the targeted tasks and training harness.

If post-training is programming with data, software engineering (SE) expertise should help
write the program. Decades of SE research and practices have produced knowledge and tools for specifying, modifying, and verifying software.
We argue that this knowledge should shape
what coding LLMs learn, not merely evaluate their outputs or establish safeguards around
them. Prior systems such as EM-Assist and RefactoringMirror combine LLMs with program
analysis and IDE tools to support individual refactorings~\cite{emassist,refactoringmirror}.
\system{} applies established SE techniques to construct verifiable training data for
improving the model itself, with refactoring as the organizing capability. Refactoring
preconditions and program analysis identify eligible examples, PyCharm and Ruff construct
reference transformations, and structural and regression tests that determine which
teacher-generated trajectories qualify for training. Building this data presents challenges. For example, repository histories can intertwine refactorings with other changes, while LLM-generated
transformations require independent verification. We therefore began with seven bounded,
tool-verifiable refactoring types, predominantly renames. We report the data construction
process, empirical outcomes, and engineering lessons as an industrial case study rather
than proposing a new training algorithm.

Our results indicate that SE-guided training data can improve a targeted
engineering capability and transfer to related tasks that are not explicitly covered during training.
\system{} constructed 5,704 executable tasks from \onboardingAccepted{} Python
repositories and retained 4,572 verified trajectories. For C1, incremental fine-tuning of
an existing 92B coding LLM on these trajectories, comprising 72.51 million tokens, i.e., 0.46\% of
its preceding coding mixture, improved performance by \textbf{\rqoneRefactorbenchAbsoluteGain{}
percentage points on RefactorBench and \rqonePromaxAbsoluteGain{} points on SWE-Bench
ProMax}. The gains extended to related SWE issue-resolution tasks (C2), with improvements of
\rqtwoVerifiedAbsoluteGain{} points on SWE-Bench Verified and \rqtwoProAbsoluteGain{}
points on SWE-Bench Pro. Although we
collected the training trajectories using OpenCode, the fine-tuned coding LLM achieved higher average scores
under all four evaluated harnesses (C3), including a statistically significant improvement
under Mini-SWE-Agent on SWE-Bench Verified (\rqthreeVerifiedMiniPValue{}).
Transfer remained task-dependent:
performance on greenfield repository generation improved under two harnesses but regressed under two
others. These findings demonstrate the potential of training interventions focused on
a bounded capability,
while underscoring the challenges of consistent transfer across tasks and harnesses.

As coding LLMs increasingly contribute to software development,
SE must shape how these models are developed.
Debates over agentic and AI-native SE largely focus
on how engineers will work with these systems~\cite{cao2026agentic,hassan2026se3}.
Our industrial case study demonstrates a complementary role: we used
established SE knowledge to select engineering capabilities, construct
verified training data, and evaluate model improvements. Building
\system{} also revealed practical challenges, including losses during
training-data construction, the constraints verification imposes on
capability coverage, and the need to assess improvements and regressions
across workloads and agent harnesses. Building on this experience, we
call on the SE community to contribute directly to coding LLM training,
develop engineering methods for model development and maintenance,
and systematically define the capabilities these models should acquire,
drawing on established knowledge in SWEBOK~\cite{swebok2024}.

\section{Industrial Background and Motivation}
\label{sec:motivation}

\subsection{Industrial Background}
\label{sec:industrial-background}

\smallskip \noindent \textbf{Industrial setting.} The study was conducted at our large international technology
company with multiple product lines and heterogeneous software systems that differ in
languages, architectures, and development conventions. Our company pretrained a
92B-parameter coding LLM and has refined it through several rounds of post-training. The model now
serves more than 100,000 internal developers through different coding agents. Our team maintains and updates this coding LLM, whose capabilities must serve diverse
engineering workloads rather than individual products or repositories.

\smallskip \noindent \textbf{Maintaining and updating the coding LLM.} The coding LLM evolves
through post-training, which combines data targeting different capabilities into a shared
training mixture. At this scale, post-training requires substantial computing resources and the allocation of a finite data budget across competing objectives. Full post-training runs therefore
follow scheduled cycles rather than being conducted as individual research experiments. Industrial
model-development reports describe the same constraint~\cite{mai_thinking_1,rajbahadur2026llm}.
Between cycles, research teams develop candidate training datasets and evaluate their effectiveness through
incremental fine-tuning of an existing checkpoint. Before proposing these datasets for inclusion in the next
post-training mixture, our team assesses the three challenges introduced in
Section~\ref{sec:introduction}: whether verified training data improves the targeted
capability (C1), whether gains transfer to related tasks without unacceptable regressions in
other capabilities (C2), and whether improvements persist across agent harnesses (C3). These
criteria guide our evaluation in Section~\ref{sec:design}.

\subsection{Motivating Study: Tool-Guided Refactoring Across Repositories}
\label{sec:motivation-study}

\smallskip \noindent \textbf{Motivation.} A recent industrial report described substantial
maintainability gains from tool-guided refactoring, achieved through considerable engineering
effort. Tornhill~\cite{codescenescale} reported increasing the \textit{Code Health} of a 300,000-line codebase from 5.6 to 10 with CodeScene, which assesses more than 25 maintainability factors
on a scale from 1 to 10~\cite{codescenehealthdefinition}. The effort took three weeks, involved 2,903
commits, approximately US\$4,000 in inference costs, an evolving project-specific
refactoring playbook, and specialized behavioral verification. In our setting, we cannot
build comparable guidance and verification infrastructure for every repository, nor commit
such resources to each one. We therefore investigated whether a frontier coding agent could achieve
meaningful improvements across repositories with a common refactoring procedure and a fixed
budget, without project-specific skills or playbooks.

\smallskip \noindent \textbf{Approach.} We selected five repositories from the DeepSWE
long-horizon task set~\cite{datacurve2026deepswe} to evaluate a common procedure across
different codebases rather than a single customized environment. We ran Codex with GPT-5.6
Sol at \texttt{xhigh} reasoning effort. Guided by CodeScene feedback, the agent identified
refactoring candidates, generated characterization tests when needed, and retained a change
only when project tests passed and \textit{Code Health} improved. The workflow ran
autonomously under a budget of US\$1,500 until no acceptable change remained or its
allocated budget was exhausted.

\begin{table*}[!t]
\caption{CodeScene-guided refactoring across five projects. Code Health gains remained relatively modest despite substantial time and model costs. Size is small ($<$5k LOC), medium (5k--10k), or large ($>$10k). Final health is the score of the repository after the last accepted refactoring, and $\Delta$ is the relative change in percentage. Acc.\ and Rej.\ are the numbers of accepted and rejected refactoring iterations. Refactoring time is the agent's total runtime during the refactoring process.}
\label{tab:motivation}
\centering\footnotesize
\setlength{\tabcolsep}{2pt}
\begin{tabular*}{\linewidth}{@{\extracolsep{\fill}}lrlrrrrrrrrl@{}}
\toprule
& & & \multicolumn{3}{c}{Code Health} & \multicolumn{2}{c}{Iterations} & \multicolumn{3}{c}{Cost} & \\
\cmidrule(lr){4-6}\cmidrule(lr){7-8}\cmidrule(lr){9-11}
Project & LOC & Size & Initial & Final & $\Delta$ (\%) & Acc. & Rej. & \shortstack{Tokens\\(M)} & USD & \shortstack{Time\\(h:min)} & Status \\
\midrule
Arktype    & 32,617 & Large  & 8.89 & 8.99 & 1.12  & 19 & 3 & 383.58 & 268.44 & 19:31 & Budget exhausted \\
Participle &  7,859 & Medium & 8.04 & 8.46 & 5.22  & 24 & 3 & 668.07 & 454.24 & 28:29 & Budget exhausted \\
ABS        &  6,596 & Medium & 6.93 & 7.43 & 7.22  & 20 & 2 & 591.20 & 425.81 & 28:43 & Budget exhausted \\
Dateutil   &  5,725 & Medium & 3.88 & 5.64 & 45.36 & 20 & 7 & 263.07 & 165.29 & 10:09 & Finished \\
Aiomonitor &  2,653 & Small  & 8.19 & 8.67 & 5.86  & 10 & 0 & 274.72 & 186.22 & 12:01 & Finished \\
\midrule
Total      & 55,450 & ---    & ---  & ---  & ---   & 93 & 15 & 2,180.64 & 1,500.00 & 98:53 & --- \\
\bottomrule
\end{tabular*}
\end{table*}

\smallskip \noindent \textbf{Results.} Three of the five runs exhausted their budgets before
completing, despite substantial time and inference expenditure. Together, the runs consumed
nearly 99 hours, 2.18 billion tokens, and US\$1,500. The unfinished runs improved
\textit{Code Health} by 1.12\% to 7.22\%, corresponding to absolute gains of 0.10 to 0.50 points on the 10-point scale. Among
the two completed runs, Aiomonitor improved by 5.86\% (0.48 points). We observed a maximum
absolute improvement of 1.76 \textit{Code Health} points in Dateutil (3.88 to 5.64); although
this represented a 45.36\% relative gain, its final score remained below the other
repositories' initial scores (Table~\ref{tab:motivation}).

Our results suggest that tool-guided refactoring alone may yield limited
improvements under our industrial constraints. Despite substantial agent effort, maintainability
improvements were modest and uneven. Our shared coding LLM serves developers across
heterogeneous repositories, where we cannot rely on open-ended agent budgets or repository-specific
refactoring playbooks. Together with the uneven refactoring performance in
Figure~\ref{fig:frontier-radar}, these observations motivated us to investigate whether SE
knowledge could strengthen the coding LLM's underlying refactoring capability through
training. Section~\ref{sec:approach} presents \system{}, which constructs the verified
training data needed to improve this capability.

\section{RefactorForge: From SE Knowledge to Verified Training Data}
\label{sec:approach}
We constructed \system{} to leverage established SE knowledge to generate verified training data for improving the refactoring capability of our coding LLM. \system{} first constructs executable tasks from controlled refactoring transformations in real repositories. Each task has a known objective and independently verifiable outcome. A teacher coding agent then attempts each task without access to its reference patch or hidden grading tests. We retain trajectories only whose final solutions pass the specified checks. Each retained trajectory records the task instruction, tool calls, repository observations, and code edits. We then use these successful trajectories to train our coding LLM through supervised fine-tuning (SFT), as described in Section~\ref{sec:setup}.

We drew on SE knowledge to guide how \system{} constructs
and verifies tasks. Specifically, we used refactoring preconditions to identify
eligible code, established engineering tools to construct reference transformations,
and structural and behavioral checks to define acceptable solutions.
Following the executable-task format established by SWE-Bench~\cite{jimenez2024swe},
\system{} uses fail-to-pass (F2P) tests that fail before the transformation and pass
afterward, and pass-to-pass (P2P) tests that remain passing in both states.
Figure~\ref{fig:pipeline} summarizes the twelve-step pipeline across four phases.
We first explain how we selected refactoring types, then present the state of
practice, SE-informed decision, and implementation for each phase.

\begin{seipregularfigure}[t]
\centering
\includegraphics[width=\columnwidth]{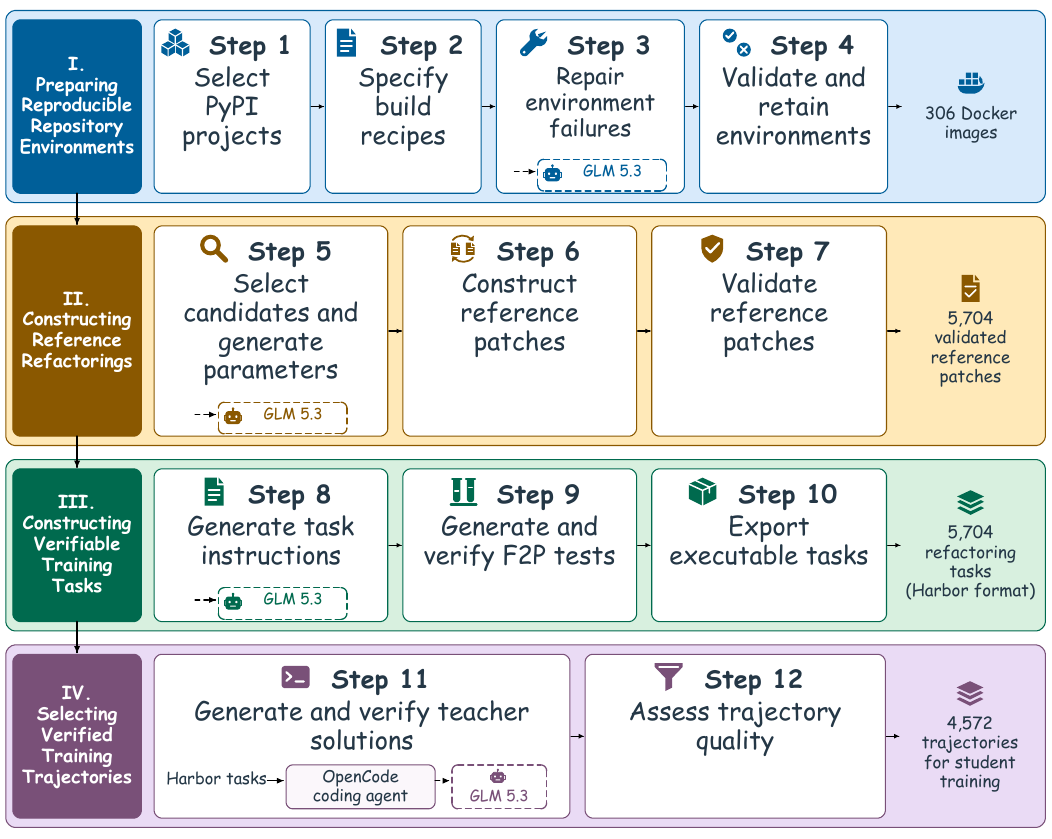}
\caption{Overview of \system{}'s twelve-step pipeline.}
\Description{Four horizontal phases run from top to bottom, with steps flowing from left to right within each phase: Preparing Reproducible Repository Environments (steps 1--4), Constructing Reference Refactorings (steps 5--7), Constructing Verifiable Training Tasks (steps 8--10), and Selecting Verified Training Trajectories (steps 11--12). Phase I selects project versions, specifies recipes, repairs failures, and validates builds before retaining recipes and 306 Docker images and recording passing project tests and baseline execution time. Black arrows connect phases and steps. Dashed arrows connect environment repair, Rename parameter proposals, instruction generation, and the OpenCode teacher agent to GLM 5.3. \texttt{Move Function} destinations are selected by rules. Phase II outputs 5,704 validated reference patches for the reported tasks. Phase III produces 5,704 refactoring tasks in Harbor format. Harbor tasks supply instructions and original project versions to the teacher. Step 11 verifies teacher solutions with F2P and P2P tests. Step 12 assesses trajectory quality, leaving 4,572 trajectories for student training.}
\label{fig:pipeline}
\end{seipregularfigure}

\subsection{Selecting Refactoring Types}
\label{sec:refactoring-types}

We selected refactoring operations that require repository-level code changes and have explicitly verifiable outcomes. Each type specifies eligibility conditions and a structural objective that the pipeline can check, allowing us to construct targeted training tasks without relying on an LLM to define the correct transformation.

\smallskip \noindent \textbf{State of Practice.} Existing coding-model training pipelines use several approaches to task construction. SWE-Gym and SWE-smith derive tasks from repository issues or synthetic faults~\cite{swegym,swesmith}, while SWE-Flow uses test-driven development to construct incremental training tasks~\cite{sweflow}. Refactoring-specific training focuses on individual operations: Palit and Sharma~\cite{palitrefactor} train models for Java Extract Method, and RefBERT studies variable-name recommendations~\cite{liu2023refbert}. Our pipeline requires executable repository-level tasks across several refactoring types, including operations that coordinate changes to definitions and references across files.

\smallskip \noindent \textbf{SE-Informed Decision.} We selected the Rename, Move, and Remove families based on empirical evidence about refactoring practices and coding-model limitations. Murphy-Hill et al.~\cite{murphyhill2009refactor} found that developers frequently rename program elements and perform move refactorings. Caivano et al.~\cite{caivano2023deadmethods} documented persistent unused methods, while Ebrahimi et al.~\cite{ebrahimi2026add} found that coding LLMs can avoid deleting code even when the intended change requires removal. These operations provide explicit training targets while requiring agents to locate definitions, trace dependencies, and coordinate changes across files.

\smallskip \noindent \textbf{Implementation.} \system{} supports seven types:
\texttt{Rename Function}, \texttt{Rename Class}, \texttt{Rename Constant},
\texttt{Rename Attribute}, \texttt{Move Function}, \texttt{Remove Unused Function}, and
\texttt{Remove Unused Import}. Table~\ref{tab:refactoring-types} lists their candidate code,
structural F2P properties, and task counts. Our pilot targets correct execution of specified
transformations rather than judging whether a refactoring improves software design. We
therefore limit coverage to operations that established tools can apply and our checks can
verify. Renames account for 96.4\% of generated tasks.
\begin{table}
\caption{Refactoring candidates, F2P test checks, and task counts. Function and class renaming account for 80.12\% of tasks.}
\label{tab:refactoring-types}
\label{tab:f2p-properties}
\centering\small\setlength{\tabcolsep}{4pt}
\setlength{\bigstrutjot}{\baselineskip}
\begin{tabular}{@{}>{\raggedright\arraybackslash}p{0.16\linewidth}>{\raggedright\arraybackslash}p{0.23\linewidth}>{\raggedright\arraybackslash}p{0.36\linewidth}>{\raggedleft\arraybackslash}p{0.17\linewidth}@{}}
\toprule
Refactoring type & Candidate code & Properties checked by generated F2P tests & Tasks (share) \\
\midrule
\texttt{Rename Function} & Top-level functions. & \multirow[t]{3}[3]{=}{\raggedright New name present in the defining module and old name absent. Where affected by the reference change, identifier uses, import bindings, and \texttt{\_\_all\_\_} exports also reflect the rename.} & 1,899 (33.29\%) \\[\baselineskip]
\addlinespace
\texttt{Rename Class} & Top-level classes. &  & 2,671 (46.83\%) \\[\baselineskip]
\addlinespace
\texttt{Rename Constant} & Module-level constants. &  & 567 (9.94\%) \\
\addlinespace
\texttt{Rename Attribute} & Class and instance attributes. & New name replaces the old name in the owning class. Affected attribute references reflect the rename. & 362 (6.35\%) \\
\addlinespace
\texttt{Remove Unused Function} & Private top-level functions with no resolved references. & Target function absent. Remaining top-level definition names match the reference outcome. & 25 (0.44\%) \\
\addlinespace
\texttt{Remove Unused Import} & Names flagged as unused in \texttt{from ... import ...} statements. & Target import bindings absent. Retained bindings and top-level definition names remain. & 46 (0.81\%) \\
\addlinespace
\texttt{Move Function} & Top-level functions moved between repository modules. & Function defined in the destination, not the source. Affected \texttt{from ... import ...} statements target its new location. & 134 (2.35\%) \\
\midrule
\textbf{Total} & & & \textbf{5,704 (100\%)} \\
\bottomrule
\end{tabular}
\end{table}

\subsection{Preparing Reproducible Repository Environments (Phase I)}
\label{sec:recipes}

We prepared reproducible repository environments and established behavioral baselines for subsequent refactoring task construction and verification
Each transformation must execute against a known project version, and \system{} must check whether previously passing tests continue to pass.

\smallskip \noindent \textbf{State of Practice.}
Constructing executable repository environments remains a bottleneck in
coding-agent training~\cite{swesmith}. Existing pipelines use LLM agents
to identify dependencies and resolve installation failures. SWE-smith
combines agent-assisted environment discovery with manual
review~\cite{swesmith}, while SWE-rebench uses execution feedback to
refine model-generated installation recipes~\cite{swerebench}.
Scale-SWE, SWE-Factory, and daVinci-Env use agents for the construction of the environment~\cite{scaleswe,swefactory,davincienv}.
Although these approaches reduce manual configuration effort, they incur inference costs without
guaranteeing successful environment construction~\cite{swefactory}. These
costs can become substantial at scale~\cite{davincienv}.

\smallskip \noindent \textbf{SE-Informed Decision.} We approached environment preparation as a problem of reproducible builds and regression testing. We designed \system{} to derive installation recipes from package metadata, apply explicit repair rules, and reserve bounded LLM assistance only for unresolved failures. We accept an environment only after it passes validation and its frozen recipe was successfully replayed. However, these requirements restrict the pool of repositories available for training-task construction, as repositories without reproducible environments are excluded.

\smallskip \noindent \textbf{Implementation.} Phase I comprises four steps.

\pipelinestep{Select PyPI projects}
\label{project-pinning}
We selected actively maintained packages from the Python Package Index (PyPI) because its structured release, dependency, and installation metadata supports automated environment construction and reproducible project-version selection. For each package, \system{} identifies the upstream repository, maps the release to a tag, and records the corresponding commit hash to fix the project version.

\pipelinestep{Specify build recipes}
\system{} extracts compatible Python versions and build and test dependencies from package metadata, optional dependency groups, requirements files, and tox configurations. It uses these to generate Dockerfiles, then attempts project installation, test collection, and test execution before requesting LLM assistance.

\pipelinestep{Repair environment failures}
\system{} first applies deterministic repairs and cached solutions to address known installation failures, unsupported pytest options and markers, missing imports, and dependency incompatibilities. If these fail, \system{} queries GLM~5.3 to propose structured repairs to dependencies, Python versions, or environment variables. The model cannot execute shell commands, modify project code, or change test commands. \system{} applies and validates each proposal and caches successful repairs for reuse. Each project receives at most six build-and-test-collection attempts, two additional rounds of dependency repair, and seven LLM repair requests. Projects with unresolved failures are rejected.

\pipelinestep{Validate and retain environments}
\label{recipe-acceptance}
\system{} accepts an environment only when the project installation succeeds,
test collection completes without errors, and two testing conditions hold:
the project has at least one passing test, and no tests fail or produce errors.
Skipped tests and expected failures are permitted but do not count as passing
tests. \system{} records the passing tests and baseline execution time, freezes
the recipe, and verifies reproducibility through replay. These passing tests
form the P2P regression suite used in later phases.

We assessed \onboardingAssessed{} projects; \onboardingBuilt{} built successfully
(\onboardingBuildRate{}), \onboardingTested{} reached test execution
(\onboardingTestExecutionRate{}), and \onboardingAccepted{} met the final acceptance
criteria (\onboardingAcceptanceRate{}). We incurred \onboardingLLMSpend{} in LLM expenditure
for environment construction, or \onboardingCentsPerAccepted{} cents per accepted recipe. The
remaining 423 projects did not satisfy all acceptance requirements.

\subsection{Constructing Reference Refactorings (Phase II)}
\label{sec:reference-refactorings}

We used established SE tools, refactoring preconditions, and transformation semantics to construct independently verifiable reference patches. While Phase I establishes reproducible repositories in which models can practice, Phase II determines which transformations they should learn.

\smallskip \noindent \textbf{State of Practice.} Prior work combines LLMs with SE tools to support refactoring execution. EM-Assist separates candidate generation, static analysis, and transformation, using an LLM, program analysis, and an IDE, respectively~\cite{emassist}. MANTRA instead uses collaborating agents to generate and repair refactored code~\cite{mantra}, while RefactoringMirror uses IntelliJ IDEA to reapply transformations detected in LLM-generated revisions~\cite{refactoringmirror}. Although these approaches support refactoring execution, constructing training data also requires reference transformations with independently checkable correctness criteria. Obtaining such transformations presents two challenges. First, mining repository histories requires separating refactorings from accompanying changes, as SWE-Refactor demonstrates~\cite{swerefactor}. Second, generating refactorings with LLMs can produce unsafe transformations that available unit tests fail to detect, as Liu et al. found~\cite{refactoringmirror}.

\smallskip \noindent \textbf{SE-Informed Decision.} We chose to generate reference patches using refactoring semantics rather than unconstrained LLM-generated changes. We defined eligibility conditions and structural transformation procedures for each refactoring type. \system{} uses program analysis to identify eligible symbols and PyCharm or Ruff to apply the specified transformations. We restricted LLM assistance to bounded parameters, such as alternative identifiers, rather than code changes. This approach avoids dependence on mined developer patches but limits coverage to transformations supported by the selected tools and checks.

\smallskip \noindent \textbf{Implementation.} Phase II comprises three steps.

\pipelinestep{Select candidates and generate parameters}
\label{candidate-detection}
\system{} uses PyCharm's program-analysis APIs to identify definitions, resolve repository-wide usages, and evaluate type-specific eligibility conditions. For example, \texttt{Remove Unused Function} targets private, undecorated top-level functions with no resolved usages. For \texttt{Remove Unused Import}, \system{} uses Ruff's unused-import diagnostics, restricted to \texttt{from ... import ...} statements. For each eligible candidate, \system{} records its source location, current name, and relevant usage metadata.

For rename operations, we require plausible alternative identifiers that \system{} could validate without allowing an LLM to modify code. \system{} uses GLM~5.3 to propose a different, meaning-preserving identifier from the symbol's name, type, file path, and surrounding code. Deterministic filters reject unchanged names, invalid identifiers, Python keywords, and changes to the original leading-underscore category. These filters also enforce PEP~8 naming conventions: \texttt{snake\_case} for functions and attributes, \texttt{CapWords} for classes, and \texttt{UPPER\_SNAKE\_CASE} for constants.

For \texttt{Move Function}, \system{} identifies plausible destinations from the distribution of resolved function usages, favoring modules that already depend on the function. It counts usages by repository module, excluding import statements, and selects the module outside the source module with the highest count. The destination must contain at least two usages and more usages than the source module. \system{} breaks ties by lexicographic file-path order and rejects destinations containing a conflicting definition or an import binding the same name to a different symbol. However, \system{} permits existing imports of the function being moved. This heuristic identifies a plausible training destination, not necessarily the architecturally optimal one. Remove operations require no additional parameters.

\pipelinestep{Construct reference patches}
\label{refactoring-application}
\system{} uses PyCharm's program-structure and refactoring APIs to apply supported rename and move transformations, and Ruff's diagnostic fixes to remove unused imports. These tools construct patches without LLM-generated code changes. For each candidate, \system{} applies the transformation to the project version pinned in Step~\ref{project-pinning}, records the patch and refactoring metadata, and restores the original repository before processing the next candidate.

Although each transformation has a bounded objective, its patch can require coordinated changes across files. Across 5,704 generated tasks, the median patch changes two files, spans four diff hunks, and involves eight lines of patch churn (lines added plus deleted). At the 95th percentile, patches change six files, span 17 hunks, and involve 56 lines of churn.

\pipelinestep{Validate reference patches}
\label{refactoring-validation}
\system{} validates each reference patch by applying it to the original repository and executing syntax and operation-specific structural checks. For renames, it checks naming conventions and incomplete reference updates. For \texttt{Move Function}, it verifies that the definition appears at the destination, disappears from the source, leaves unrelated top-level definitions unchanged, and introduces no duplicate imports. \system{} then reruns every project test that passed in Phase I and accepts the patch only if all structural checks pass and previously passing tests remain passing.

We thus obtained reference patches that passed the implemented structural checks and the project tests that had passed before the transformation.

\subsection{Constructing Verifiable Training Tasks (Phase III)}
\label{sec:task-construction}

A validated reference patch specifies what the model should change, but does not establish
how to assess an agent's solution. In Phase III, \system{} converts each reference
transformation into an executable instruction with independently verifiable correctness
criteria.

\smallskip \noindent \textbf{State of Practice.} Coding-agent benchmarks commonly use
executable F2P and P2P tests. Pipelines such as SWE-smith and R2E-Gym construct tasks
from synthetic faults or development histories~\cite{swesmith,r2egym}. For refactoring,
however, passing project tests does not establish that the requested transformation
occurred. Hong et al.~\cite{swerefactorbench} demonstrate this distinction in SWE Refactor
Bench. Across 520 migration runs, 30 preserved tested behavior but failed the migration
audit, while 252 satisfied the audit but failed behavioral tests. SWE-Refactor,
RefactorBench, and CodeTaste also combine structural checks with behavioral
validation~\cite{swerefactor,refactorbench,codetaste}.

\smallskip \noindent \textbf{SE-Informed Decision.} We used refactoring-specific correctness
obligations to construct training tasks. \system{} generates instructions that specify the
intended operation without revealing its implementation, requiring agents to locate and
modify the relevant code. To verify the resulting changes, \system{} derives structural
F2P tests from the reference transformation and reuses previously passing project tests
for regression checking.

\smallskip \noindent \textbf{Implementation.} Phase III comprises three steps.

\pipelinestep{Generate task instructions}
\label{instruction-generation}
\system{} provides GLM~5.3 with the refactoring type, symbol metadata, and validated
reference patch to generate a task instruction. Following RefactorBench's Lazy
instruction style, the model produces a brief developer-facing request that
identifies the required symbols but omits file paths and patch contents.
\system{} checks the instruction against the recorded metadata. Instructions containing unsupported details, file paths, or raw patch syntax are rejected, and a revision is requested.

\pipelinestep{Generate and verify F2P tests}
\label{test-generation}
\system{} derives structural F2P tests from each reference transformation to check
operation-specific changes to names, definitions, references, and imports. It
validates each test against both repository states, retaining it only if it fails on the original repository and passes after the reference patch is applied.
Where appropriate, \system{} also reuses project tests modified by the reference
transformation to exercise the changed interface. Table~\ref{tab:f2p-properties}
summarizes the structural properties checked for each refactoring type.

\pipelinestep{Export executable tasks}
\label{task-export}
\system{} packages each validated refactoring as an executable
Harbor~\cite{harborframework} task containing the pinned project version, Docker
environment, setup procedure, task instruction, and F2P and P2P tests.

In Phase III, \system{} produced 5,704 executable tasks across seven
refactoring types.

\subsection{Selecting Verified Training Trajectories (Phase IV)}
\label{sec:teacher-trajectories}

Executable tasks establish what coding agents must accomplish and how to verify
their solutions. In Phase IV, \system{} collects demonstrations of how a teacher
agent solves these tasks, and we use engineering-defined correctness to determine
which trajectories enter the training mixture.

\smallskip \noindent \textbf{State of Practice.} Coding-model post-training commonly
uses executable checks to select successful teacher-agent trajectories for training.
SWE-Gym, Skywork-SWE, SWE-Lego, and Open-SWE-Traces follow this approach to
collect demonstrations for repository-level
tasks~\cite{swegym,zeng2025skywork,swelego,ahmad2026openswetraces}.

\smallskip \noindent \textbf{SE-Informed Decision.} Unlike the preceding phases,
Phase IV introduces no additional SE-specific mechanism. We adopted established
post-training data collection practice, using the refactoring correctness criteria defined in
the preceding phases to select teacher trajectories. \system{} retains a
trajectory only when its final solution satisfies both the structural and
behavioral requirements.

\smallskip \noindent \textbf{Implementation.} Phase IV comprises two steps.

\pipelinestep{Generate and verify teacher solutions}
\label{sec:distillation}
We selected GLM~5.3 as the teacher model to generate solutions and execution
trajectories for the refactoring tasks. Its publicly available weights support
self-hosted inference, and its reported 95.4\% SWE-Bench Verified score under
Mini-SWE-Agent demonstrates substantial repository-level coding
capability~\cite{valsai2026swebench}.
For each executable task, \system{} runs GLM~5.3 through the
OpenCode~\cite{opencode} coding-agent harness. The teacher receives the task
instruction and original repository but neither the reference solution nor
the hidden grading tests. \system{} checks the teacher's final solution
against both F2P and P2P tests and retains the trajectory only if it passes
both. Of 5,704 tasks, 4,889 (85.71\%) yielded successful teacher trajectories.

\pipelinestep{Assess trajectory quality}
\label{sec:trajectory-quality}
\system{} further filters successful trajectories for malformed conversations,
incomplete responses, excessive repetition, garbled text, restricted terms,
non-English characters, and invalid multi-turn structure. Unlike the F2P and
P2P checks, these filters assess the trajectory rather than the final code.
After filtering, 4,572 trajectories remained for student training.

Together, these phases produce verified trajectories for training our model.

\section{Industrial Case Study and Findings}
\label{sec:design}

We used the distilled trajectories retained after verification and quality filtering in
Phase IV for refactoring-oriented SFT. We evaluated the intervention against the three
criteria our team uses when considering data for a subsequent post-training mixture
(Section~\ref{sec:industrial-background}): improvement in the targeted capability (C1),
transfer to related tasks with acceptable effects on other capabilities (C2), and
persistence of improvements across agent harnesses (C3).

\subsection{Study Design}
\label{sec:setup}

\smallskip \noindent \textbf{Training.}
Our baseline, \baselinemodel{}\footnote{Anonymized for double blind}, was our company's 92B-parameter
coding LLM (Section~\ref{sec:industrial-background}), post-trained from
a pre-trained model using a mixture spanning multiple capabilities.
While we cannot disclose the full mixture, its code-related portion
comprised 938,180 trajectories and 15.66 billion tokens.
As discussed in Section~\ref{sec:industrial-background}, we used incremental
supervised fine-tuning (SFT) to evaluate the effectiveness of new
training data between scheduled full post-training cycles.
We fine-tuned \baselinemodel{} on 4,572 verified \system{} trajectories
totaling 72.51 million tokens (0.46\% of the preceding code-related
allocation), producing \refactoringmodel{}.
These trajectories averaged 11.77 agent turns and covered seven refactoring
types across the Rename, Move, and Remove families.
Training ran for 6 epochs and 83 iterations on 12 nodes, with a
batch size of 48 sequences and a maximum sequence length of 128K tokens
(full settings in the supplementary material~\cite{refactorforge-replication}).

\smallskip \noindent \textbf{Evaluation.}
We designed our evaluation around the three challenges identified in
Section~\ref{sec:industrial-background}. For each benchmark, we report
Avg. @3, the average score across three independent runs per model, and the difference between models
in percentage points (pp). We also compare each model's highest-scoring
run using a one-sided exact McNemar test~\cite{mcnemar1947}.
Because both models attempt the same tasks, McNemar tests differences
in paired binary outcomes rather than treating success rates as
independent. We report matched-pair
odds ratios (OR), where OR greater than 1 favors \refactoringmodel{}.

\smallskip \noindent \textbf{C1: Improving refactoring capability.}
We evaluated targeted refactoring using two benchmarks:
\textbf{RefactorBench}~\cite{refactorbench} (Lazy set) and \textbf{SWE-Bench ProMax}~\cite{promax}.
The Lazy set contains 100 Python tasks with brief instructions that
require agents to infer the necessary repository changes,
while ProMax contains 170 refactoring tasks across seven
languages. We used OpenCode~\cite{opencode} for both
benchmarks, matching the harness used to collect teacher trajectories.

\smallskip \noindent \textbf{C2: Transfer to other engineering tasks.}
We evaluated transfer beyond targeted refactoring using three benchmarks:
\textbf{SWE-Bench Verified}~\cite{sweverified}, \textbf{SWE-Bench Pro}~\cite{swepro}, and
\textbf{NL2Repo-Bench}~\cite{nl2repo}. The first two assess issue resolution. SWE-Bench Verified contains 500 human-screened Python issues, while SWE-Bench Pro's public v1 set contains
731 bug-fixing and feature-implementation tasks across four
languages. SWE-Bench Pro is significantly more challenging and requires
long-horizon reasoning. NL2Repo-Bench assesses repository generation,
requiring agents to construct a complete, installable Python repository,
including its architecture, dependencies, and implementation, from a
natural-language requirements document and an empty workspace.

We also explored potential code-quality benefits by using
PyExamine~\cite{shivashankar2025pyexamine} to compare code smells
introduced by the two models when resolving the same Python
\textbf{SWE-Bench Pro} tasks. We restricted the comparison to tasks
both models resolved and patch pairs PyExamine successfully analyzed.

\smallskip \noindent \textbf{C3: Transfer across agent harnesses.}
We assessed whether improvements persist beyond OpenCode, the harness
used to generate training trajectories. We evaluated both models on
the SWE issue-resolution and greenfield benchmarks using four agent harnesses:
OpenCode~\cite{opencode}, Claude Code~\cite{claudecode},
OpenHands~\cite{wang2025openhands}, and
Mini-SWE-Agent~\cite{yang2024sweagent,minisweagentdocs}.

\subsection{Findings}
\label{sec:findings}
\smallskip \noindent
\textbf{C1: Fine-tuning on verified refactoring trajectories improves refactoring performance.}
\label{sec:c1-results}

\begin{seipregulartable}[t]
\caption{Refactoring and SWE issue-resolution results (Avg. @3).
$^{*}$ denotes $p<0.05$ (one-sided exact McNemar test).}
\label{tab:results}
\centering\footnotesize
\setlength{\tabcolsep}{3pt}
\begin{tabular}{@{}lrrrr@{}}
\toprule
& \multicolumn{2}{c}{\shortstack{Refactoring (C1)}}
& \multicolumn{2}{c}{\shortstack{SWE issue resolution (C2)}} \\
\cmidrule(lr){2-3}\cmidrule(l){4-5}
& \shortstack{SWE-Bench\\ProMax}
& \shortstack{RefactorBench\\(Lazy)}
& \shortstack{SWE-Bench\\Verified}
& \shortstack{SWE-Bench\\Pro} \\
\midrule
\baselinemodel{}    & 51.18 & 43.00 & 62.00 & 48.97 \\
\refactoringmodel{} & 58.24 & 61.00 & 66.97 & 50.20 \\
\textbf{Improvement (pp)}
& \textbf{+7.06}
& \textbf{+18.00}$^{*}$
& \textbf{+4.97}
& \textbf{+1.23} \\
\bottomrule
\end{tabular}
\end{seipregulartable}

\textit{Refact-92B outperforms Baseline-92B by 18.00 percentage
points (pp) on RefactorBench and 7.06 pp on SWE-Bench ProMax.}
Success rates increased from 43.00\% to 61.00\% on RefactorBench, and from 51.18\% to 58.24\% on ProMax (Table~\ref{tab:results}). The ProMax improvement
is notable because the benchmark evaluates large-scale
refactoring across seven programming languages, with tasks modifying
11.4 files on average, whereas our training data came exclusively
from Python repositories.

\begin{seipregularfigure}[!t]
\centering
\includegraphics[width=0.80\columnwidth]{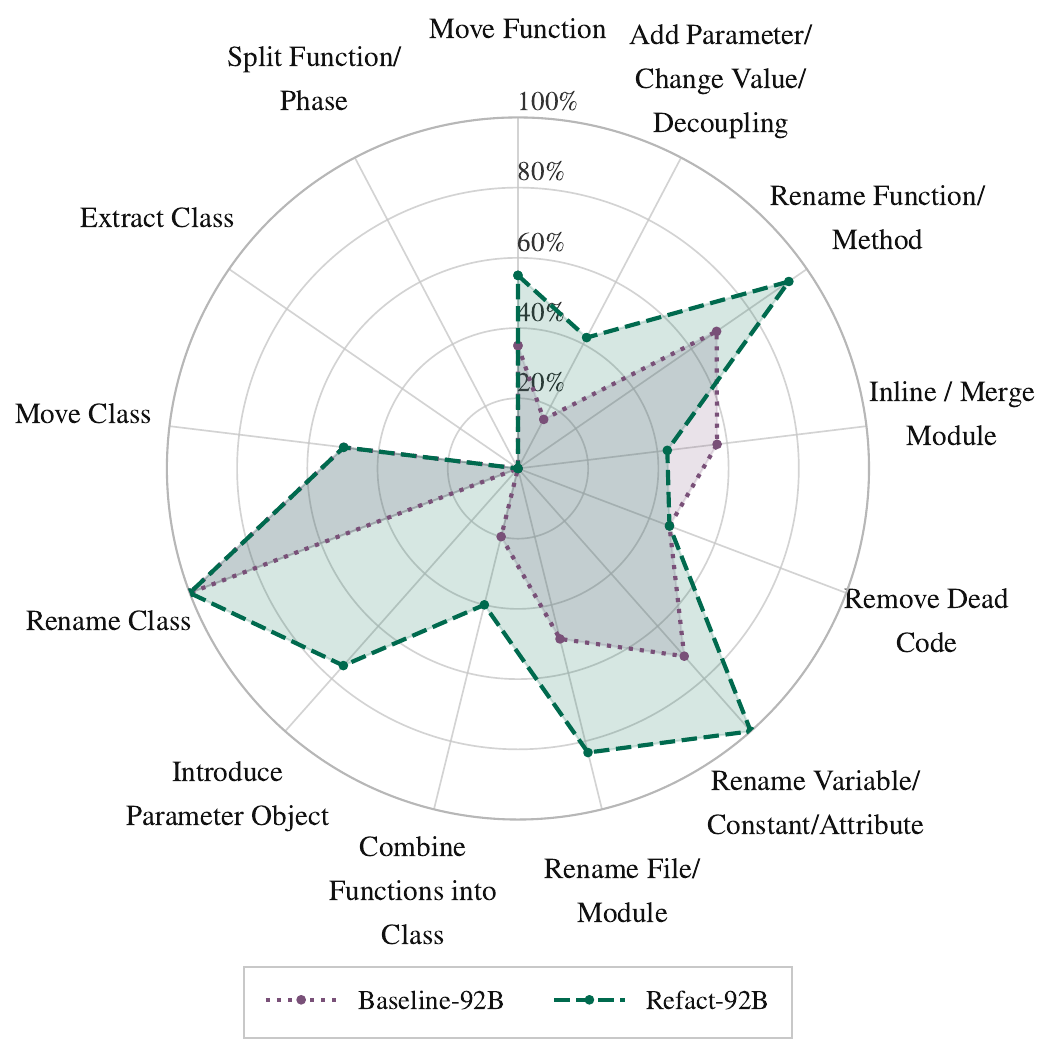}
\caption{RefactorBench Lazy success rates by refactoring category for \baselinemodel{} and
\refactoringmodel{}. \refactoringmodel{} reduced unevenness across refactoring operations.}
\label{fig:c1-radar}
\end{seipregularfigure}

\textit{Refact-92B outperforms Baseline-92B in seven of thirteen
refactoring categories and successfully completed tasks in eleven,
despite training on only seven refactoring types.}
Renames accounted for 96.4\% of our training tasks, yet improvements
extended to other operations. For example, success on Introduce Parameter Object increased from 0\% to 75.0\% (Figure~\ref{fig:c1-radar}). The number of categories with success rates below
50\% also decreased from seven to five, producing a more even
performance profile. Refact-92B's overall success rate of 61\%
approached GPT-5.2's 62\% and exceeded GLM-5.1's 51\% in our
motivating evaluation (Figure~1). Refact-92B also succeeded in
11 of 13 categories, compared with 10 for GPT-5.2 and 12 for
GLM-5.1. These results show how a bounded, capability-focused
training intervention can strengthen performance across refactoring operations beyond those explicitly trained.

\smallskip \noindent
\textbf{C2: Performance gains extend beyond refactoring to other SE tasks.}
\label{sec:c2-results}

\textit{Refact-92B outperforms Baseline-92B by 4.97 pp on
SWE-Bench Verified and 1.2 pp on SWE-Bench Pro, extending gains
beyond the refactoring tasks used for training.}
Under OpenCode, SWE-Bench Verified increased from 62.00\% to
66.97\%, while SWE-Bench Pro increased from 48.97\% to 50.20\%
(Table~\ref{tab:results}). SWE-Bench Pro is demanding, evaluating
long-horizon bug fixing and feature implementation across four
programming languages. These improvements extend to related
SE tasks that require changes to existing
repositories rather than explicitly specified refactorings.

\textit{Refact-92B also outperforms Baseline-92B on greenfield repository
generation under two of four agent harnesses, despite greenfield development
not being a training target.}
On NL2Repo-Bench, Refact-92B improved over Baseline-92B under
Claude Code, from 26.94\% to 28.88\%, and OpenHands, from
28.12\% to 31.00\%. It declined under Mini-SWE-Agent, from
27.65\% to 24.20\%, and OpenCode, from 27.96\% to 25.12\%.
NL2Repo-Bench presents
a different and demanding workload: agents must construct complete,
installable repositories, including their architecture, dependencies,
and implementation, from natural-language requirements and an
empty workspace. Unlike the consistent improvements on related
SWE issue-resolution tasks, gains on this less directly related workload
depended on the harness.

\textit{Refact-92B introduced 6.9\% fewer detected code smells than
Baseline-92B when resolving the same software issues.}
Of the 131 Python SWE-Bench Pro tasks both models resolved,
PyExamine successfully analyzed 111 patch pairs. It detected
54 introduced code smells in Refact-92B's patches, compared with
58 in Baseline-92B's patches. This exploratory result extends
our evaluation beyond task completion to properties of the
generated code.

\smallskip \noindent
\textbf{C3: Performance improvements persist across different agent harnesses.}
\label{sec:c3-results}

\begin{seipregularfigure}[t]
\centering
\includegraphics[width=\columnwidth]{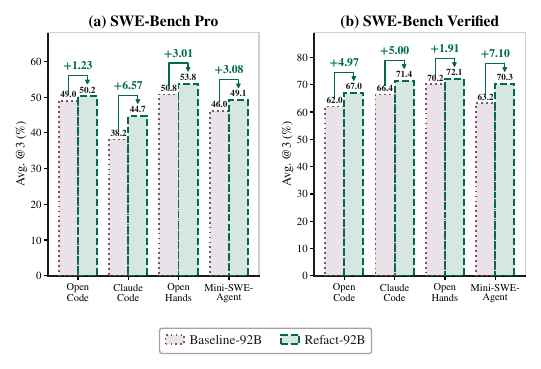}
\caption{Average scores from three independent runs across four agent harnesses.
Panel (a) shows SWE-Bench Pro and panel (b) SWE-Bench Verified. Purple dotted bars
show \baselinemodel{} and green dashed bars \refactoringmodel{}. Bar labels give
average scores, and arrows give absolute improvements in percentage points.}
\label{fig:cross-harness}
\end{seipregularfigure}

\textit{Refact-92B outperformed Baseline-92B in all eight
benchmark-harness combinations across four coding-agent
harnesses.}
On SWE-Bench Pro, improvements ranged from 1.23 to 6.57 pp,
while SWE-Bench Verified improvements ranged from 1.91 to
7.10 pp (Figure~\ref{fig:cross-harness}). Refact-92B consistently outperformed
Baseline-92B under OpenCode, Claude Code, OpenHands, and
Mini-SWE-Agent, demonstrating that the improvements persisted
across different agent execution environments.

\textit{The largest improvements on both SWE issue-resolution benchmarks
occurred under harnesses absent from training-data collection.}
On SWE-Bench Pro, Claude Code produced the largest gain of
6.57 pp, from 38.16\% to 44.73\%. On SWE-Bench Verified,
Mini-SWE-Agent produced the largest gain of 7.10 pp. We collected
all training trajectories using OpenCode, yet the largest
improvements emerged under harnesses with different tools,
prompts, and execution conventions. These results demonstrate
that the gains extended beyond familiarity with the harness used for training data collection.

\section{Lessons learned}
\label{sec:lessons}

In this section, we present the lessons learned from developing
\system{} to construct capability-focused training data and evaluating its effects on a coding LLM designed to support diverse workloads at industrial scale.

\subsection{Lessons for Building Capability-Focused Training Data}
\label{sec:lessons-data}

\smallskip \noindent \textbf{L1. Account for losses at each stage of training-data construction.}
Generating executable tasks does not guarantee usable training data.
Of the 729 projects we assessed, 614 built successfully, but only
306 (42.0\%) satisfied our environment acceptance criteria.
These environments yielded 5,704 executable tasks, of which the
teacher agent successfully solved 4,889. Subsequent trajectory-quality
filtering left 4,572 trajectories for training. Each stage imposes
distinct requirements, from reproducible environments and correct
solutions to suitable trajectories for training. Teams must account for these reductions when planning training-data generation and allocating resources.

\smallskip \noindent \textbf{L2. Plan capability coverage together with verification.}
The refactoring operations \system{} could include in training depended
on its ability to construct and verify their transformations.
We selected seven refactoring types based on established
practices~\cite{murphyhill2009refactor}. For instance, renames accounted for 96.4\% of generated tasks
(Table~\ref{tab:refactoring-types}). This concentration also reflects their verifiability: PyCharm can identify eligible symbols and update affected references, while structural and regression tests can check whether the requested transformation occurred and previously passing tests remain passing. As SWE Refactor Bench demonstrates, verifying that a transformation occurred and that it preserved behavior requires distinct
checks~\cite{swerefactor}. The availability of such checks therefore shapes which engineering capabilities we can reliably train our models on.

\texttt{Extract Class} illustrates this challenge. Neither
\baselinemodel{} nor \refactoringmodel{} solved the \texttt{Extract Class} tasks
in RefactorBench (Figure~\ref{fig:c1-radar}). However, we were not able to generate training tasks for this operation. Although we can verify a prescribed \texttt{Extract Class} transformation, deciding which members belong in the extracted class requires design judgment that our structural checks and existing tests cannot capture. Expanding capability coverage therefore requires not only additional task generators but also verification methods that capture more complex engineering objectives.
important role for SE research.

\subsection{Evaluating Capability Improvements in Coding LLMs}
\label{sec:lessons-evaluation}

\smallskip \noindent \textbf{L3. Evaluate capability improvements against regression gates.}
Post-training an existing coding LLM requires the same regression-testing
discipline as maintaining conventional software~\cite{rajbahadur2026llm}. \refactoringmodel{} improved refactoring and related SWE issue-resolution performance but regressed on greenfield repository generation under
two of four harnesses. Refactoring and issue resolution share operations such as locating code, tracing dependencies, and preserving behavior, whereas greenfield generation requires additional capabilities our training data does not explicitly target. Our incremental fine-tuning experiment revealed these trade-offs before integration into the full post-training mixture, highlighting the importance of regression gates
when evaluating model updates.

\smallskip \noindent \textbf{L4. Evaluate improvements across the agent harnesses used in practice.}
Benchmark improvements under one harness do not establish that those gains persist across different agent configurations~\cite{thangarajah2026dcas}. \refactoringmodel{} improved SWE-Bench Pro under all four evaluated
harnesses, but gains ranged from 1.2 pp under OpenCode to 6.57 pp under Claude Code (Figure~\ref{fig:cross-harness}). The largest improvements on both SWE issue-resolution benchmarks occurred outside OpenCode, which we used to collect training trajectories. Evaluating only the collection harness would have missed these differences.

\section{SE Must Shape Coding LLM Development:
A Research Agenda}
\label{sec:implications}

Recent advances in coding LLMs have intensified questions about the
future of SE research. Salsabil et al.~\cite{salsabil2026survives}
revisited 35 LLM-based techniques from ICSE 2026 and found that a
newer model using a single prompt outperformed 13, with mixed results
for another nine. Importantly, techniques providing domain-specific
knowledge and external feedback proved more resilient, highlighting
the enduring value of engineering expertise. Meanwhile, ongoing
discussions question how increasingly capable coding agents will
reshape software development and whether the roles of developers and
SE expertise will diminish~\cite{cao2026agentic,hassan2026se3,
welsh2023end,yellin2023premature,brynjolfsson2025canaries}. We see a
different opportunity. Our experience demonstrates how SE knowledge
can directly improve the coding LLMs driving these changes. Rather
than diminishing, we argue that our field's role must expand to
developing these models themselves. We propose three calls to action
for SE research.

\smallskip \noindent
\textbf{Evolve coding LLMs into software engineering LLMs.} If
post-training is programming with data~\cite{rajbahadur2026llm}, software
engineering must help write that program. The training data used to develop and
refine coding LLMs shapes the engineering behaviors they acquire. Our
field has decades of knowledge about how software should be designed,
constructed, tested, and maintained. Research such as SWE-smith
~\cite{yang2025swesmith} already demonstrates the value of contributing
training data directly to model development. \system{} extends this
direction by using established refactoring knowledge to determine
what the model practices and what constitutes a correct solution. This
reflects the value of domain knowledge identified by Salsabil et
al.~\cite{salsabil2026survives}. SE researchers must bring their
expertise into training-data construction, from selecting engineering
operations to generating executable tasks and developing verification
methods. These contributions can move us from coding LLMs toward
software engineering LLMs, improving the engineering capabilities of
the models themselves rather than only the systems built around them.

\smallskip \noindent
\textbf{Apply SE to coding LLM development and maintenance.}
Developing coding LLMs requires more than generating training data.
Prior work characterizes industrial post-training as brownfield
maintenance, where teams update existing checkpoints under finite
budgets while preserving established capabilities~\cite{rajbahadur2026llm}.
This introduces familiar SE challenges throughout model development,
from training-data construction and integration to regression testing
and release decisions. SE researchers must develop regression-aware
methods that predict how training-data changes affect other
capabilities without repeatedly running expensive benchmark suites.
Early work on data-mixture performance prediction and proxy metrics
provides promising foundations~\cite{ye2025datamixing,patel2026forecasting}.
These methods should complement automated regression testing across
model updates, workloads, and agent harnesses. Beyond regression
testing, SE research must also advance methods for validating
training data, ensuring reproducibility, tracing the effects of data
changes, and managing model integration and releases. Together, these
opportunities call for applying established SE principles throughout
the model-development lifecycle.

\smallskip \noindent
\textbf{Systematically define what coding LLMs should learn.}
Individual training interventions cannot address the breadth of
capabilities required for SE. Our field needs a systematic account of
what coding LLMs should learn and how to verify that they have learned
it. SWEBOK provides a foundation by organizing established knowledge
across requirements, design, construction, testing, maintenance, and
other areas~\cite{swebok2024}. SE researchers must translate this
knowledge into explicit engineering capabilities, their constituent
operations, appropriate training tasks, and correctness criteria.
These specifications must also identify relationships among
capabilities and guide evaluations of acquisition and transfer.
\system{} provides an initial example for refactoring, while its
verification challenges demonstrate the work needed for more complex
engineering activities. Such a research agenda would connect our
field's accumulated knowledge to the capabilities we develop in
coding LLMs.

Much of our community's current research examines how to use coding
LLMs in SE tasks~\cite{steenhoek2025closing}, how to evaluate their
performance~\cite{jimenez2024swebench}, and where their capabilities
remain insufficient~\cite{ding2025primevul}. These contributions
remain essential, but our field has a larger role to play. We must
move beyond studying and adapting to increasingly capable coding LLMs
and actively contribute to their development. As coding LLMs
increasingly engineer software, SE must help engineer the models
themselves.

\label{sec:conclusion}
\section{Conclusion}

Moving from coding LLMs to software engineering LLMs requires our field to shape both what models learn and how we develop them. Our industrial case study demonstrates one way to achieve this. Through \system{}, we applied established SE knowledge, tools, and verification practices to construct training data that improved refactoring and related issue-resolution performance across agent harnesses. We evaluated one 92B model and seven bounded refactoring types, predominantly renames. Constructing verified training data, fine-tuning large models, and evaluating their improvements require substantial engineering effort, computational resources, and financial investment. Our findings therefore provide an initial demonstration rather than evidence of generality across models and capabilities. We call on the SE community to systematically define the capabilities these models should acquire and apply its expertise throughout model training, evaluation, development, and maintenance.

\bibliographystyle{seip/IEEEtran}
\bibliography{references,motivation-references,harness-references,refactoring-selection-references}
\end{document}